\pdfoutput=1

\documentclass[	DIV=classic,%
paper=a4,%
fontsize=11pt,%
onecolumn]{scrartcl}

\usepackage{comment}
\usepackage{fancyhdr}						
\usepackage[margin=1in]{geometry}

\usepackage[title]{appendix}

\usepackage{amsmath}
\usepackage{amsfonts}
\usepackage{amssymb}
\usepackage{graphicx}
\usepackage{epstopdf}
\usepackage{color}
\usepackage[caption=false]{subfig}
\usepackage{latexsym}
\usepackage{listings}
\usepackage{verbatim}
\usepackage{mathtools}
\usepackage{ctable}
\usepackage[multiple]{footmisc}

\usepackage{booktabs,tabularx}
\usepackage[group-separator={,}, group-minimum-digits=4, input-decimal-markers={.}]{siunitx}
\usepackage{lipsum}
\usepackage{url}
\usepackage{abbrevs}

\newcommand{\etal}{\textit{et al. }}

\providecommand{\keywords}[1]{\textbf{\textit{Keywords: }} #1}

\title{Modelling a novel Coronavirus (COVID-19):\\ A stochastic  SEIR-HCD approach, with real-time parameter estimation \& \\ forecasting for Scotland}

\author{
	Jonathan Wells \footnotemark[2] \footnotemark[1] 
	\and Chris Robertson \footnotemark[2] 
	\and Vincent Marmara \footnotemark[4] 
	\and Alan Yeung \footnotemark[5]
	\and Adam Kleczkowski \footnotemark[2]
}

\begin{document}
	\maketitle
	\renewcommand{\thefootnote}{\fnsymbol{footnote}}
	\footnotetext[2]{Department of Mathematics \& Statistics, University of Strathclyde, Glasgow, G1 1XH, UK.}
	\footnotetext[1]{Email: jonathan.wells@strath.ac.uk}
	\footnotetext[4]{Faculty of Economics, Management \& Accountancy, University of Malta, Msida, MSD 2080, Malta.}
	\footnotetext[5]{Health Protection Scotland, Public Health Scotland, Glasgow, G2 6QE, UK.}
	\renewcommand{\thefootnote}{\arabic{footnote}}
	\begin{abstract}
		
		Faced with the 2020 SARS-CoV2 epidemic, public health officials have been seeking models that could be used to predict not only the number of new cases but also the levels of hospitalisation, critical care and deaths. In this paper we present a stochastic compartmental model capable of real-time monitoring and forecasting of the pandemic incorporating multiple streams of real-world data, reported cases, testing intensity, deaths, hospitalisations and critical care occupancy. Model parameters are estimated via a Bayesian particle filtering technique. The model successfully tracks the key variables (reported cases, critical care and deaths) throughout the two waves (March-June and September-November 2020) of the COVID-19 outbreak in Scotland. The model hospitalisation predictions in Summer 2020 are consistently lower than the recorded data, but consistent with the change to the reporting criteria by the Health Protection Scotland on 15th September. Most parameter estimates were constant over the two waves, but the infection rate and consequently the reproductive number decrease in the later stages of the first wave and increase again from July 2020. The death rates are initially high but decrease over Summer 2020 before rising again in November. The model can also be used to provide short-term predictions. We show that the 2-week predictability is very good for the period from March to June 2020, even at early stages of the pandemic. The model has been slower to pick up the increase in the case numbers in September 2020 but forecasting improves again in the later stages of the epidemic.
	\end{abstract}
	

	\pagestyle{myheadings}
	\markboth{A. KLECZKOWSKI, J. WELLS, C. ROBERTSON \etal }{{\scshape Modelling Covid-19: A stochastic SEIR-HCD approach.}}
	
	\section{Introduction - COVID-19}
	The first cases of the newly discovered coronavirus COVID-19 (also SARS-CoV-2, and previously 2019-nCoV) were reported in Wuhan, the capital of the Hubei Province in China, on \num{29}\textsuperscript{th} December 2019, and all four were linked to the Huanan Seafood Wholesale Market \cite{li2020early}. This virus is a novel form of coronavirus-infected pneumonia that generally presents itself in the form of a fever, tiredness and a dry cough \cite{WHO_Covid_19_FAQ}. However, there are also asymptomatic cases where infected people show no symptoms at all. Since the discovery of these initial cases, the virus has subsequently spread across the globe, with the first two cases in the United Kingdom (UK) and first in Scotland being reported on \num{31}\textsuperscript{st} January and the \num{1}\textsuperscript{st} March 2020, respectively \cite{Worldometer_data_updates}.
	
	As the virus spread across the globe, public health authorities have been requesting models that could be used to predict not only the number of new cases but also the levels of hospitalisation, critical care and deaths. The ability to provide an early warning has been crucial to their decision-making process, as the increase in these numbers could lead to the break-down of the health system.
	
	In response to the request from Public Health Scotland which manages the devolved health system in this part of the United Kingdom, we have adapted an existing stochastic SEIR (Susceptible-Exposed-Infected-Recovered) epidemic model \cite{Marmara_2014} to allow the real-time monitoring and forecasting of the spread of COVID-19 in the Scottish context. In doing so, additional category compartments have been added for hospitalised patients, critical care patients and deaths. The approach adopted was first presented by Ong \etal \cite{ong2010real}, and applied to model the spread of the H1N1-2009 Influenza A pandemic when it reached Singapore. Here, they collected data for new influenza A cases from a network of general practitioner and family doctor clinics and used these streams of data to re-fit the model parameters daily, enabling them to successfully monitor and predict the spread of the epidemic across Singapore. An adaptation of this model has more recently been presented by Marmara \etal to successfully estimate the force of infection for the 2009-2010 influenza A epidemic on Malta \cite{Marmara_2014}, however, as far as we are aware, this approach has not yet been applied to COVID-19.
	
	\subsection*{COVID-19 models}
	
	Many established epidemiological models have now been adapted and redeployed to forecast the COVID-19 pandemic. Previously, these same models have successfully captured the spread of other diseases --- including ebola, influenza, dengue and the 2002 SARS-CoV-1 and 2012 MERS-CoV coronavirus outbreaks. Typically, models and parameter estimation approaches use either one or only a restricted number of data streams (typically cases or death records). They also assume that all or most parameters are constant in time, except for the infection rate. Our model differs from most currently existing approaches in that it incorporates five streams of data to estimate time-varying coefficients.
	
	Within the current COVID-19 literature, there exist multiple variations of the classic Kermack McKendrick SEIR models \cite{kermack1991contributions}, generally applied using constant parameters, such as the model presented in \cite{NG_Basic_SEIRS_2020}. A useful summary of these models can be found in \cite{sameni2020mathematical}, where models are presented in the context of the COVID-19 pandemic.
	
	One of the most referenced mathematical models for COVID-19 is the Imperial College SEIR (Susceptible-Exposed-Infected-Recovered) compartmental model \cite{ferguson2020report}, which forecasts the effects of non-pharmaceutical interventions on virus spread. This model has now been widely used to justify government policies and interactions both in the UK and across the globe, with the model parameters are carefully informed based on prior observations.  
	
	Among the variations upon classical models, \cite{Liu_2020March11, magal_webb_SIRU_Other_countries} present SIR (Susceptible-Infected-Recovered) models that split the infected population into three separate compartments; reported symptomatic, unreported symptomatic and asymptomatic. The model then assumes that the reported symptomatic cases are isolated immediately, and so have no influence on the spread of the virus. They use this to model the total number of cases, whilst illustrating the effects of asymptomatic transmission and timely government restrictions, for China, South Korea, France and Germany. The majority of model parameters are fixed for all time; however, they do incorporate a time-dependent transmission rate, by fitting an exponential curve a single stream of real-world data describing the cumulative number of reported cases.  
	
	For France, \cite{Roques_2020} estimates an infection fatality ratio of \num{0.8}\%, using a mechanistic statistical variation of a SIR model, to work backwards from official hospital deaths data. Here, they also estimate a symptomatic-to-asymptomatic ratio of \num{1}:\num{8}. The mortality rate of infected patients is fitted to a single stream of real-world deaths data, using maximum likelihood estimation in MATLAB software. The other model parameters are estimated using Bayesian Markov chain Monte Carlo (MCMC) methods and are fixed constant for all time. 
	
	A deterministic SEIR model with additional compartments for hospitalisations, the asymptomatic cases, and deaths, is presented in \cite{EIKENBERRY2020293}. This paper also considers both a fixed and time-varying infectious contact rate parameter - where an exponential curve is used to capture the transmission rate. However, most parameters are constant for all time and are based upon informed estimates from the literature, with no fitting to real-world data.
	
	Examples of further deterministic and statistical COVID-19 SEIR models incorporating hospitalisation, intensive care and deaths can be found in \cite{TuiteE497, neto2020compartmentalized,Chowdhury}. However, again, all of these models use parameter estimates that are constant for all time, based upon figures within the literature. 
	
	In \cite{ROOSA_Feb_2020}, three different phenomenological models are presented and used to produce $10$--$14$ day forecasts of the expected number of new COVID-19 cases for Wuhan China. The three models used are the generalised logistic growth model, generalised Richards model and a sub-epidemic wave model \cite{VIBOUD_2016,original_richards_1959,Chowell_2019}, all of which have been combined with a parametric bootstrap approach to estimate model parameters using data for real-world cumulative cases in Wuhan, China \cite{CHOWELL_2017}. In all three cases, these models use a single stream of real-world data (confirmed cases), to fit the model parameters and subsequently forecast the cumulative daily number of cases.
	
	A statistical data analysis approach is used in \cite{Murray2020} to produce a short-term forecast of {COVID-19} deaths, by fitting to existing virus data. Here, they fit a mixed-effects non-linear regression model to the real-world cumulative deaths data for a specific location. This approach is combined with a micro-system model that is fitted to hospitalisation data, to forecast health service needs (Intensive care beds, etc.) from the deaths forecast. However, the predicted number of deaths seems exceptionally high, under this approach --- with a forecast of \num{1584737} deaths in the United States alone, due to the first wave. (This figure is larger than the current death toll worldwide). Again, only a  single stream of real-world data, for deaths, is used to fit model parameters to the outbreak. 
	
	A deterministic variation upon an SEIRD model (SEIR plus deaths), with time-dependent parameters is described in \cite[Section VI.B]{sameni2020mathematical}, where an extended Kalman filter is used to fit the model parameters to the real-world observations. However, the model framework is presented purely for illustration purposes in this paper and to our knowledge has not been applied to fit parameters to real-world COVID-19 data. 
	
	A deterministic SEIHRD (SEIR with hospitalisation and death) model is presented in \cite{ivorra2020mathematical}, defines some model parameters based upon time-dependent curves that have been fitted to the real-world data, however, as these curves are fixed shapes they cannot account for discontinuous changes in parameter values as observed for $R_t$ following the lockdown. 
	
	A Bayesian Monte Carlo approach to modelling COVID-19 is presented in \cite{Dana_BRAM_COD}, and successfully used to predict the spread of the virus in Brazil. However, model parameters, such as importation rates are set to fixed distributions for all time, so these do not evolve, as the virus dynamic evolves in time. 
	
	A deterministic compartmental model that incorporates critical care patients and hospitalisations, with a time (and lockdown) dependent infectious contact rate parameter, is presented in \cite{prabhakaran2020spread}. Here, other model parameters are fixed, based upon examples in the literature, and the initial infection rate and number infected are fitted to real-world mortality and observed new cases data.
	
	In \cite{zhang2020estimation}, real-time data on the number of cases aboard the Diamond Princess cruise ship is fed into a bootstrap resampling algorithm, and combined with \textsf{R} software packages  ``projections'' and ``get\_R'' to estimate $R_t$, and obtain a forecast of new cases for the next ten days. However, again, this model uses a single stream of real-world data only. 
	
	\subsection*{Our approach}
	
	In this paper we build on the model and parameter estimation technique of Marmara et al \cite{Marmara_2014} to incorporate multiple data streams and apply it to the surveillance data for Scotland between March and November 2020. We first discuss the available data, and point out some peculiarities in the way this data has collected. As our parameter estimation technique is based on weakly informative priors, we subsequently present the rationale underlying our choice of parameter values. We discuss the model and our implementation of the particle filtering method, before presenting the results. We particularly address the following questions: (i) \textit{Can the model capture the dynamics represented by the five streams of data (reported cases, reporting intensity, hospitalisation and critical care occupancy and deaths)?} (ii) \textit{How do the parameters change in time, reflecting the changes in the processes underlying the virus transmission and its public health implications?} (iii) \textit{Can the model be used for short-term prediction (14 days) for the key public health indicators (hospitalisation, critical care, deaths)?} We finish with the future directions and recommendations.
	
	\section{Data}
	
	In this section we describe the data used in the study and the values of key parameters that guided our choice of weakly informative priors.
	
	\subsection*{COVID-19 surveillance data: Scotland}
	
	Our paper focuses on Scotland, using daily surveillance data to demonstrate the capabilities of the model; both as a tool for analysing the spread of the virus and its impact on local public health systems (both retrospectively), and as a system for short-term forecasting.  Real-time data for the total number of reported cases globally, as well as for individual countries is widely available online \cite{Worldometer_data_updates,WHO_Covid_19_Situation_Map,JH_Github_resources}.  However, they typically are not characterised by the granularity that is essential for making detailed predictions. In response to the previous pandemic threats (particularly in the wake of the H1N1 outbreak), Scottish authorities have established a highly effective surveillance system. Comprehensive daily-updated data for Scotland, can be found at \cite{Corona_infection_survey_ONS,Covid_gov_UK,Scotland_Covid19_trends}, including hospitalisations, critical care patient numbers and deaths. 
	
\begin{figure}[!h]
		\centering           
		\includegraphics[width=\textwidth]{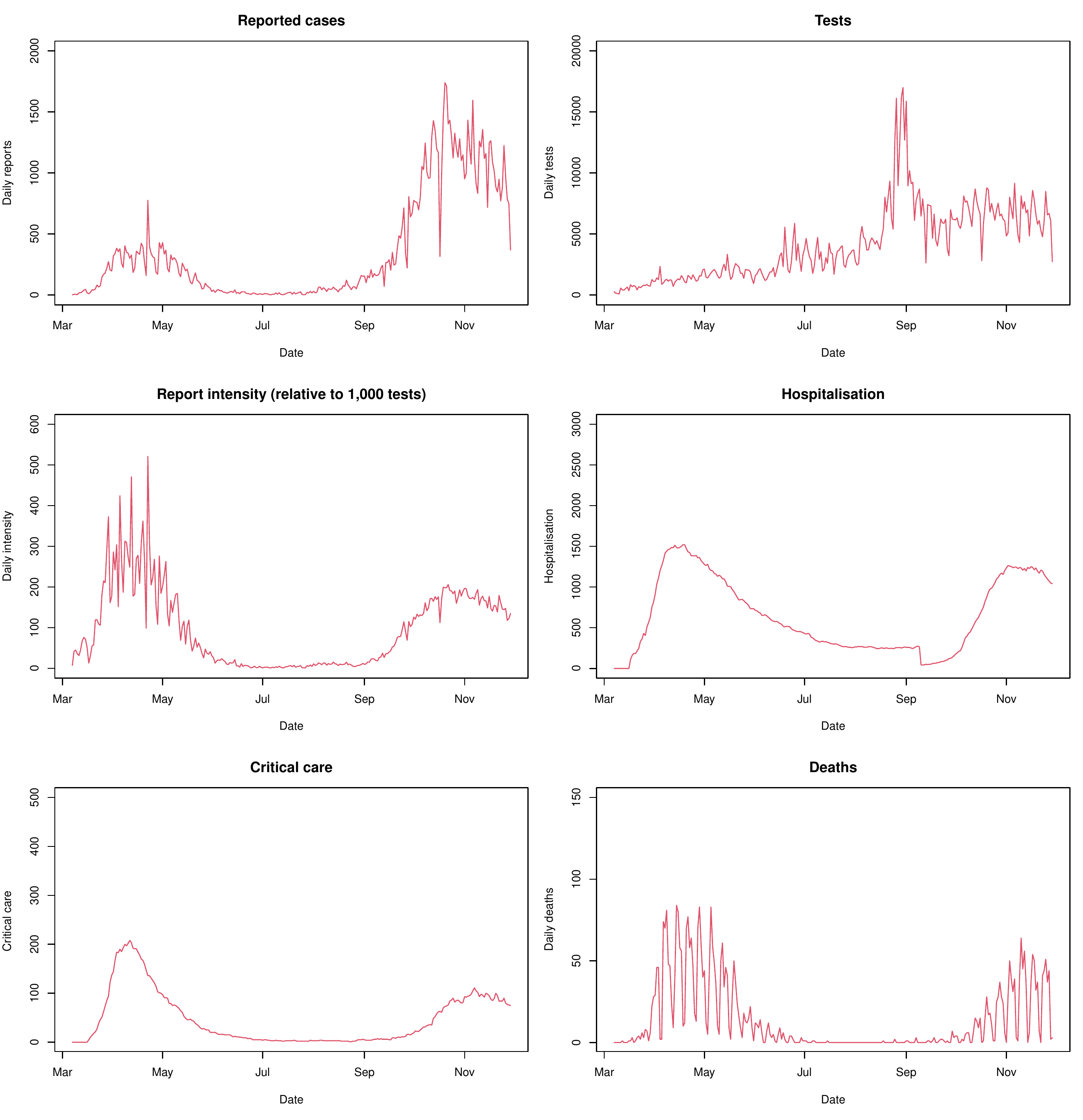} 
		\caption{Real world COVID-19 surveillance data for Scotland, from \num{8}\textsuperscript{th} March--\num{30}\textsuperscript{th} November. This data is publicly available online at \cite{Scotland_Covid19_trends}. (a) Reported cases; (b) number of tests; (c) case intensity (the number of cases scaled by the number of tests with the baseline of {1,000} tests per day); (d) current number of hospitalised patients; (e) current number of patients in critical care, and (d) daily deaths.}
		\label{Figure_1}
	\end{figure}

	The surveillance data, detailing the evolution of the COVID-19 pandemic within Scotland, is illustrated in Figure \ref{Figure_1} and includes: reported cases, tests carried out, report intensity, the number currently hospitalised, the number currently in critical care wards and deaths, for the period from 8\textsuperscript{th} March to 30\textsuperscript{th} November 2020. The data includes the first wave, from March to June 2020, which was successfully controlled using a nationwide lockdown alongside social distancing interventions. The second wave, starting in September and still ongoing at the time of the writing, is being combated using similar techniques. 
	
	One of the critical difficulties in modelling epidemic data is the relative low-efficiency and time-changing reporting of cases which the models need to incorporate. Initially, in Scotland, testing for COVID-19 took place solely within hospitals, and hence, initial reports only describe the most serious of symptomatic cases, Figure \ref{Figure_1}(a). In reality, it is expected that there were significantly more, less severe/asymptomatic cases present within the wider Scottish population, that were not captured within the surveillance data. 
	
	Testing capacity significantly increased throughout 2020, Figure \ref{Figure_1}(b), beginning at around \num{100}--\num{1000} tests during the first wave, and ramping up in testing to around \num{6000}--\num{8000} tests per day from August onwards. In July, this testing data expanded from purely hospital lab test results, to include results from the government's home testing scheme, drive-through test centres and mobile testing units, amongst others. As a result, whilst the second wave of cases (September onwards) may look more prominent than the first wave, based purely upon daily case reports, the actual numbers are probably comparable. In our model, we incorporate the number of tests by scaling the number of reported cases by the number of tests so that it is equivalent to \num{1000} tests per each day, Figure \ref{Figure_1}(c).
	
	Not only the number of tests carried out has changed over the pandemic course, but on  \num{15}\textsuperscript{th} September there was a significant change to how the number of currently hospitalised COVID-19 patients was reported, Figure \ref{Figure_1}(d). Before the \num{15}\textsuperscript{th} September, the data included people who had previously tested positive for COVID-19 but who were now in hospital for another reason as well as people who were long term hospital patients, but had contracted COVID-19 within a hospital, then remained there after recovery, due to the pre-existing condition they were staying in hospital to have treated. Subsequently, the currently in hospital/critical care data, from \num{15}\textsuperscript{th} September onwards, is purely COVID-19 admissions who had tested positive in the last 28 days (with these people subsequently removed from the data after \num{28} days). This change is evident in the hospitalisation plot found in Figure \ref{Figure_1}(d), where there is a sudden drop around the \num{15}\textsuperscript{th} September. Note also that the hospitalisation data were not available until 17th March 2020 when it jumped to 112.
	
	In the Scottish surveillance data \cite{Scotland_Covid19_trends} the number of currently hospitalised patients represents the total of both those in hospital and those in critical care, cf.\ Figure \ref{Figure_1}(d) and (e). 
	
	The surveillance data for deaths, Figure \ref{Figure_1}(f), also clearly shows the two waves of COVID-19 cases sweeping across Scotland. As this data represents deaths by the date of reporting, it is also evident that there is a strong weekly signal, whereby the number of deaths reported at weekends is far smaller than the number of deaths reported Monday-Friday.

	\subsection*{Virus characteristics} 
	In the parameter estimation we used weakly informative priors, to reflect the state of knowledge when the first cases were discovered.
	
	In the absence of control measures, the COVID-19 virus has proven highly contagious. The virus is primarily spread through respiratory droplets, and human to human contact, although has also been shown to be present within blood and urine. 
	
	The basic reproductive number, $R_0$, quantifies the average number of new infections caused by a single infected individual in an entirely susceptible population. When the reproductive number is below $1$, an outbreak is shrinking, whereas, when it is bigger than $1$, the outbreak is growing larger. A systematic review of the existing literature \cite{Lui_Review} has suggested that the COVID-19 $R_0$ has a mean value of \num{3.28} and a median of \num{2.71}, whilst another review carried out a meta-analysis of 29 papers, and suggested a pooled $R_0$ estimate of \num{3.32} ($95$\% confidence interval (CI): \num{2.81}, \num{3.82}) \cite{Alimohamadi_Review_meta_analysis}. All of these estimated are notably higher than initial World Health Organisation (WHO) estimates of between $2$ and $2.5$ \cite{WHO_R_Estimate_Sit_rep_46}. 
	
	The reproductive number varies with time as an outbreak evolves, due to human interventions and reduced availability of susceptible hosts. Thus, the time dependant, effective reproductive number, denoted $R_t$, has received significant attention within the media, as governments around the world endeavour to reduce its value to below $1$. The value of $R_t$  also varies from country to country. It is affected by interventions such as country-wide lockdowns, closures of schools, and restrictions on hospitality venues such as bars and restaurants.
	
	When an individual becomes infected, there is a time lag, between them contracting the virus, and subsequently showing symptoms. This is termed the incubation period (or latent period), and its mean value for COVID-19 is estimated at \num{5}-\num{6} days by the WHO, although can last up to \num{14} days \cite{WHO_Incubation_Sit_rep_73}. Other sources present similar numbers, for example, \cite{backer2020incubation} gives a mean incubation period of \num{6.4} days (CI: \num{2.1}, \num{11.1} ). In what follows, we will assume that individuals are not infectious during the incubation period, and thus cannot pass the virus onto others. These individuals are included in the ``Exposed'' class within the model presented here. 
	
	The period for which an individual remains infectious will vary from case to case and the severity of their infection.The median infectious period has been reported as \num{9.5} days (Interquartile range: \num{6}–\num{11} ) \cite{ling2020persistence}. It has also been suggested that patients become infectious around \num{1}-\num{3} days before symptom onset \cite{wei2020presymptomatic}, with peak infectiousness and viral load occurring at the time of symptom onset \cite{He_Covid_temporal_dynamics, to2020temporal}.
	
	For the original Wuhan outbreak, the transmission rate, describing the average number of people an infectious person spreads the disease to per day, has been estimated to be  \num{1.94}\si{\per day} (CI: \num{1.25}, \num{6.71}) \cite{Read2020}. 
	
	It is currently unknown whether recovered individuals will have permanent immunity to the virus. However, in serological testing of French hospital staff whom previously tested positive for mild COVID-19 infections, virus neutralising antibodies were found in virtually all cases from $13$ days after the onset of their symptoms \cite{Kremer2020}. This implies that temporary immunity to future infection is created within infected individuals, even for mild cases of the virus. Studies into previous human coronaviruses, suggest virus suppressing antibodies generally last around 2 years for SARS-CoV-1, implying an infected person could become susceptible again roughly 3 years after initial infection \cite{wu2007duration,tse2020current}. We assume that immunity lasts longer than the time-scales being modelled despite  a number of laboratory-confirmed reports of people succumbing to reinfection with COVID-19 \cite{To_reinfection_August_2020}.
	
	Amongst those infected with the virus, there exists a significant proportion of asymptomatic carriers, who develop no symptoms at all. As a result, many of these people do not even know that they are infected. The true extent of this asymptomatic transmission is still very uncertain, however the literature suggests \num{6} -- \num{41}\% of all infections may in fact be asymptomatic, and a meta-analysis of these studies recently suggested that the proportion of asymptomatic cases is \num{16}\% (CI: \num{12}, \num{20}) \cite{byambasuren2020estimating}. A similar review put this value at \num{25.9}\% ( CI: \num{18.8}, \num{33.1}) \cite{Koh_what_do_we_know_transmission}, i.e. approximately one asymptomatic person for every three infected people that show symptoms. In the model, we do not explicitly model the asymptomatic class and include these individuals in the unobserved cases.
	
	For many, the virus can prove fatal, and the case fatality ratio for China has been estimated as \num{1.38}\%  (CI: \num{1.23}, \num{1.53}). This percentage appears to increase with age. A Bayesian network analysis estimated that the infection fatality rate (IFR) lies within the range \num{0.3} - \num{0.5}\% \cite{neil2020bayesian}. Whilst a more recent WHO report estimates that the average IFR worldwide is approximately \num{0.5} -- \num{1}\% \cite{WHO_Mortality_estimate_Aug2020}
	
	The hospitalisation rate has also been shown to vary with age, and has been estimated to range from \num{1.04}\% (CI: \num{0.62}, \num{2.13}) for those aged \num{20} -- \num{29}, and up to \num{18.4}\% (CI: \num{11.0}, \num{37.6}) for those aged \num{80} and above \cite{verity2020estimates}. In a systematic review of the literature \cite{Cook_park2020systematic}, the average time from symptom onset to hospitalisation was estimated to be \num{3.3} \si{days} (CI: \num{2.7}, \num{4.0}).
	
	For the UK population, appropriate hospitalisation rates are presented in \cite{ferguson2020report}. Here, the non-age specific hospitalisation rate is estimated to be \num{4.4}\% of all infections, based upon the age profile of the UK population. Similarly, the mortality rate of those in critical care estimated to be \num{50}\%. They also estimate the mean duration of hospitalisation to be \num{8} -- \num{10.4} \si{days}, whilst the average critical care stay is \num{10} \si{days}. Whilst a report by the Centre for Evidence-Based Medicine suggests that around \num{0.12}\% of new hospital admissions require intensive care \cite{CEBM_ICU_RATE}.
	
	\section{Mathematical model} 
	
	\subsection*{Stochastic SEIR-HCD equations}
	
	Our approach uses a modification of a stochastic SEIR compartmental model used by Ong et al \cite{ong2010real} and Marmara et al \cite{Marmara_2014} adapted specifically for the real time monitoring and forecasting of COVID-19 spread and its public health implications. 
	
	The host human population is divided into seven compartments, corresponding so the various clinical stages of COVID-19 infection: ``Susceptible'' whom are at risk of infection; ``Exposed'' who are infected, but not yet infectious; ``Infectious'' who are both infected and infectious; ``Hospitalised'' currently receiving hospital treatment; ``Critical care''  who are both hospitalised and receiving critical care intervention; ``Dead''; and finally those who have successfully ``Recovered''. The model is based on discrete-time equations using \num{24}\si{hr} increments, with the total number of people contained within each of these seven groups, at time $t$, being denoted by $S_t$, $E_t$, $I_t$, $H_t$, $C_t$, $R_t$ and $D_t$, respectively. For purpose of reporting, $H_t$ and $C_t$ are included within the ``Hospitalised'' class. 
	
	\begin{figure}[!h]
		\centering
		\includegraphics[width=0.8\textwidth, trim= 0cm 0cm 0cm 0cm,clip]{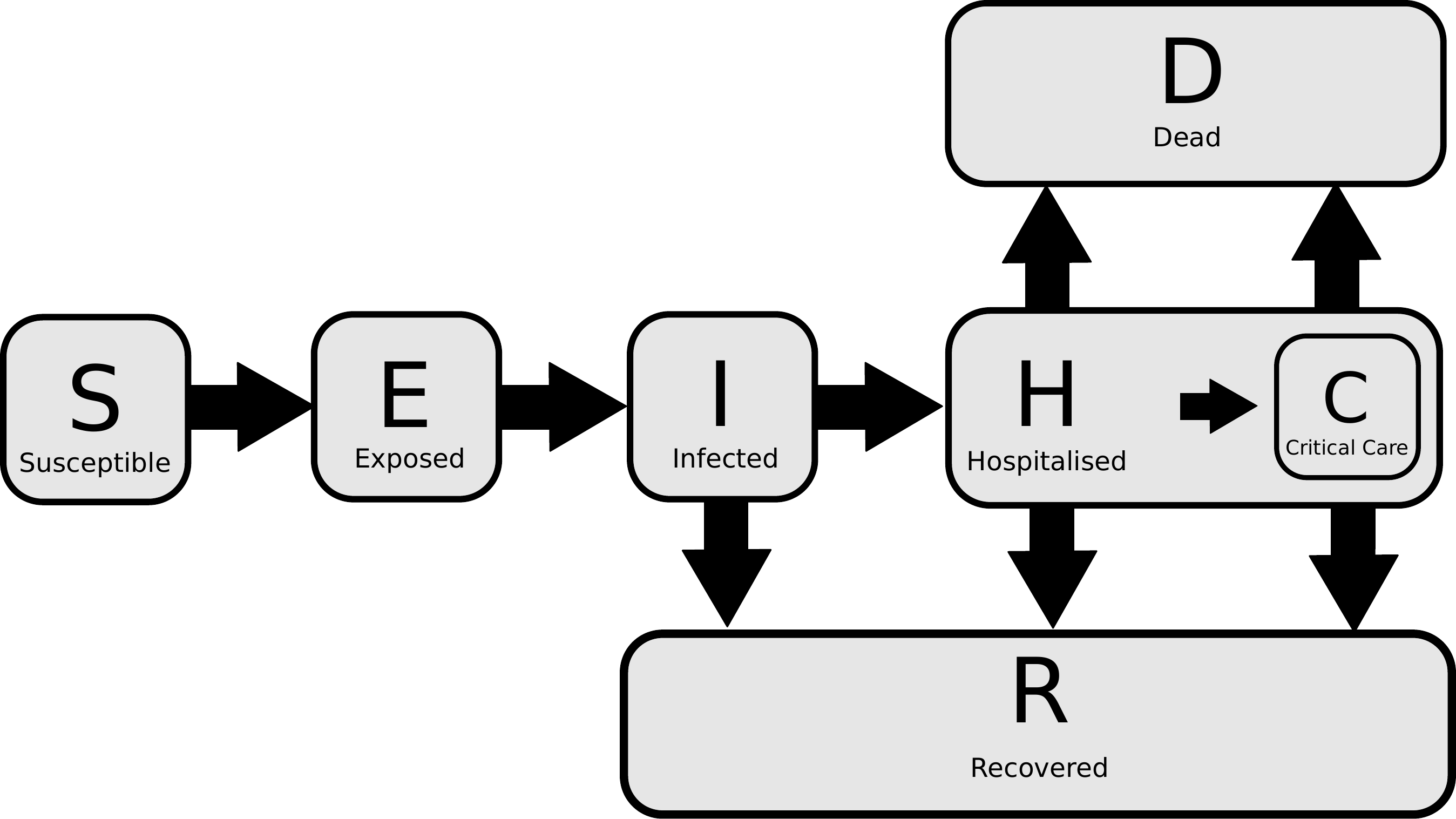} 
		\caption{SEIR-HCD model flow-chart, illustrating the model's structure. This is the assumed clinical pathway of {COVID-19} infection.}
		\label{Model_flow_chart}
	\end{figure}
	
	It is assumed that the clinical pathway of {COVID-19} infection is as illustrated in Figure \ref{Model_flow_chart}: Susceptible people become exposed to the virus at a rate of $\beta$ \si{day ^{-1}} (the transmission rate), then become infectious after a period of $\alpha$ \si{days} (termed the latent period), and subsequently move to the infected class. Infected people then either recover, after remaining infectious for $\tau$ \si{days} (the infectious period), or move into the hospitalised class. If a person is hospitalised, the model assumes the infection can take three paths; they either die, recover, or move to critical care. Under this approach, it is assumed that all deaths take place within hospital or critical care, as there is no pathway from infectious class directly to the dead class. It is also assumed, that once recovered (or dead), a person will remain immune (or dead) for the subsequent time. In addition, it is assumed that the human population is approximately homogeneous; an assumption that may not hold true if super-infectors or super-connected individuals are present; more importantly, the model is not age-structured. The total population is assumed constant and equal to $N$ for all time. 
	
	The model uses a set of seven difference equations:
	\begin{align} 
	S_t   &= S_{t-1}  - A_t           \nonumber \\  \nonumber \\
	E_t   &= E_{t-1} + A_t - B_t          \nonumber \\  \nonumber \\
	I_t   &= I_{t-1} + B_t - P^H_t  - \bar{R}^I_t            \nonumber \\ \nonumber \\
	H_t   &= H_{t-1} + P^H_t - P^{CC}_t - \bar{R}^H_t - \bar{D}^H_t   \nonumber \\ \nonumber \\
	C_t   &= C_{t-1} + P^{CC}_t - \bar{D}^{CC}_t - \bar{R}^{CC}_t     \nonumber \\ \nonumber \\
	R_t   &= R_{t-1} + \bar{R}^I_t  + \bar{R}^H_t + \bar{R}^{CC}_t           \nonumber \\ \nonumber \\
	D_t   &= D_{t-1} +  \bar{D}^H_t + \bar{D}^{CC}_t \, , \label{Gov_Eqn}
	\end{align}
	
	where $A_t$, $B_t$ and $\bar{R}^{I/H/CC}_t$ represent the number of newly infected, newly exposed and newly recovered people within the population, respectively, at time $t$.  The superscript on the newly recovered terms, $\bar{R}^{I/H/CC}_t$, simply denotes the class from which the people have recovered, e.g. $\bar{R}^{I}_t$ is the number of people from the infected compartment that recover at time $t$, whilst $H$ and $CC$ superscripts denote the same from hospital and critical care. The number of new patients in hospital and critical care are denoted $P^{H}$ and $P^{CC}_t$, whilst $\bar{D}^{H}$ and $\bar{D}^{CC}_t$ represents the number of hospital and critical care deaths at time $t$. 
	
	Stochasticity is incorporated in two ways. Firstly, the transition terms are assumed to follow binomial distributions given by:
	\begin{align} 
	A_t   &\sim \underbrace{ \mbox{Bin} \Bigl( S_{t-1}, \, 1- \exp \Bigl [   - \frac{(\epsilon - \beta I_{t-1} ) }{N}  \Bigr]     \Bigr) }_{ \mbox{Newly infected }} \nonumber \\  \nonumber \\ 
	B_t   &\sim  \underbrace{ \mbox{Bin} \Bigl( E_{t-1}, \, 1- \exp \Bigl [   - \frac{1 }{\alpha}  \Bigr]     \Bigr) }_{ \mbox{Newly infectious}} \nonumber \\  \nonumber \\   
	\bar{R}^I_t   &\sim  \underbrace{ \mbox{Bin} \Bigl( I_{t-1}, \, 1- \exp  \Bigl [   - \frac{1 }{\tau}  \Bigr]     \Bigr) }_{ \mbox{Newly recovered (from infected)}} \nonumber \\  \nonumber \\    
	\bar{R}^{H}_t &\sim \underbrace{ \mbox{Bin} \Bigl( H_{t-1}, \, 1- \exp  \Bigl [   - \frac{1 }{\lambda_H}  \Bigr]     \Bigr) }_{\mbox{Newly recovered (from hospital)}} \nonumber \\  \nonumber \\  
	\bar{R}^{CC}_t &\sim \underbrace{ \mbox{Bin} \Bigl( C_{t-1}, \, 1- \exp  \Bigl [   - \frac{1 }{\lambda_{CC}}  \Bigr]     \Bigr), }_{ \mbox{Newly recovered (from critical care)}} \label{ABX_binomials} 
	\end{align} 
	where the parameters $\tau$ and $\alpha $ are the infectious and latent periods respectively; $ \beta  $ and $\epsilon $ represent the transmission rate within and the virus importation rate into the host population (due to residents returning from abroad). Parameters $\lambda_H$ and $\lambda_{CC}$ then represent the average duration COVID-19 patients remain in hospital and critical care, respectively. The number of new hospital and critical care patients, and corresponding number of these patients dying at time $t$  are also assumed to follow binomial distributions, given by:  
	\begin{align}
	P^H_t   &\sim \underbrace{ \mbox{Bin} \Bigl( I_{t-1},  \kappa_{H} \left( 1- \exp  \Bigl [   - \frac{1 }{\lambda_{C}}  \Bigr] \right) \Bigr) }_{\mbox{New hospitalisations}} \nonumber \\  \nonumber \\ 
	P^{CC}_t  &\sim \underbrace{ \mbox{Bin} \Bigl( H_{t-1},  \kappa_{CC}  \Bigr) }_{\mbox{New critical care interventions}}  \nonumber \\  \nonumber \\ 
	\bar{D}^{H}_t &\sim \underbrace{ \mbox{Bin} \Bigl( H_{t-1},  \mu_{H}  \Bigr) }_{ \mbox{New deaths (from hospital)} }  \nonumber \\  \nonumber \\ 
	\bar{D}^{CC}_t &\sim \underbrace{ \mbox{Bin} \Bigl( C_{t-1},  \mu_{CC}  \Bigr). }_{\mbox{New deaths (from critical care)}} \label{PZ_Binomials}
	\end{align}
	Here, $\lambda_C$ is the average time from case to hospitalisation, $\kappa_{H}$ is the probability an infected individual requires hospitalization, whilst $\kappa_{CC}$ is the probability a hospitalized patient requires critical care intervention. The mortality rates of hospital and critical care patients are denoted $\mu_H$ and $\mu_{CC}$, respectively. 
	
	\subsection*{Observation process}
	
	In order to related the dynamic variables listed above to data, we introduce a stochastic observation process, based on the negative binomial distributions. Although Marmara et al \cite{Marmara_2014} uses a Poisson distribution with the mean (and variance) proportional to $I_t$ to represent the number of reported cases, we found significant overdispersion in the COVID-19 outbreak data. The negative binomial distribution provides sufficient flexibility to account for this feature, with the overdispersion being estimated from the data.
	
	The relationship between the reported number of cases, $\hat{I}^{\mbox{rep}}_t$ and the number of infected individuals $I_t$, takes the following form
	
	\begin{align}
	\hat{I}^{\mbox{rep}}_t &\sim \mbox{NegBin}\Big( 
	\Delta _{d(t)} \left( Z_t I_t + \phi \right) , \omega_I \Big)
	\label{Observ_I}
	\end{align}
	
	where the first parameter of the negative binomial distribution function represents the mean and the second one, $\omega_I$, the dispersion parameter. The background reporting rate, $\phi$, takes into account the number of false positive COVID-19 test results that might appear within the real-world data. $Z_t$ represents the data stream of daily number of tests, scaled to 1,000 tests per day. Thus, $\hat{I}^{\mbox{rep}}_t$ can be interpreted as the number of reported positive tests if exactly 1,000 tests were carried out on any day. 
	
	The daily reporting correction parameter for each week day, $\Delta _{d(t)}$ accounts for two processes. Firstly, it represents the proportion of those who sought medical help and therefore were tested, as opposed to asymptomatic individuals or those with light symptoms. Secondly, it represents weekday variation, so that $\Delta _{1}$ accounts for the probability an infected individual seeks a virus test on a Monday, $\Delta _{2}$ the same probability for a Tuesday, and so forth until $\Delta _{7}$ for Sundays. These are important here, as there is a clear trend in the real-world testing data that indicates people are more likely to test positive for COVID-19 towards the end of the week from Friday-Sunday, and less likely to seek/receive a positive test on Monday-Tuesdays. This results in a weekly oscillating signal within the data documenting the number of new daily infections. 
	
	A similar parameter, denoted $\Delta^D _{d(t)}$, is introduced to capture the corresponding weekly variation in reported deaths. This is required, as there is a clear pattern within the real world data, whereby there is a significant drop in the number of deaths reported at weekends compared to the numbers reported during Monday-Friday. This is due to the fact deaths generally tend not to be registered at weekends, and so weekend deaths figures can often be around \num{10}\% of those reported on weekdays. Thus, the number of deaths reported is given by
	
	\begin{align}
	\hat{D}^{\mbox{rep}}_t &\sim \mbox{NegBin}\Big( \Delta^D _{d(t)} D_t , \omega_D \Big).
	\label{Observ_D}
	\end{align}
	
	Finally, the hospitalisation and critical care rates are given by
	
	\begin{align}
	\hat{H}^{\mbox{rep}}_t &\sim \mbox{NegBin}\Big( H_t , \omega_H \Big)
	\label{Observ_I}\\
	\hat{CC}^{\mbox{rep}}_t &\sim \mbox{NegBin}\Big( CC_t , \omega_{CC} \Big),
	\label{Observ_CC}
	\end{align}
	where $\omega_{D}$, $\omega_{H}$ and $\omega_{CC}$ are the dispersion factors for the overdispersion in the deaths, hospitalisation and critical care data, respectively.

	\subsection*{Model priors}
	We use weakly informative priors with values and shapes obtained from the literature. The priors are summarised in Table \ref{TABLE:Model_parameters}. Here, $\mbox{T.Normal}(x,\,y)$; $[a,b]$ denotes a truncated standard normal distribution, with mean $x$, standard deviation $y$, truncated between end points $a$ and $b$. 
	
	\begin{table}[h!]
		\centering
		\renewcommand{\arraystretch}{1.5} 
		\begin{tabularx}{\textwidth}{@{}lXrr@{}}
			\toprule
			\multicolumn{1}{@{}l}{Parameter}&\multicolumn{1}{l}{Description} &\multicolumn{1}{r}{Estimated value} &\multicolumn{1}{r}{Prior distribution} \\
			\midrule
			$ \beta  $     &  Transmission rate            &          \SI{2}{day ^{-1}}  &  $ \mbox{Gamma} ( 2 \times 0.5 ,\, 0.5 )$ \\
			$  \epsilon $  &  Importation rate   &   \SI{0}{day ^{-1}} & ---  \\
			$  \tau   $    &  Infectious period  &   \SI{4}{days}  &  $ \mbox{Gamma} (4 \times 5 ,\, 5 )$    \\
			$  \alpha $    &  Latent period      &   \SI{4}{days} &  $ \mbox{Gamma} (4 \times 5,\, 5 )$     \\
			$  \phi   $    &  Background rate    &   \num{0} &  --- \\
			$\lambda_C$     &  Avg. duration from case to hospital  &   \SI{5}{days}   &  $ |\mbox{Normal} ( 10,\, 2)| $  \\
			$\lambda_H$     &  Avg. duration hospitalised  &   \SI{10}{days}   &  $ |\mbox{Normal} ( 10,\, 2)| $  \\
			$\lambda_{CC}$  &  Avg. duration critical care  &   \SI{10}{days}  &   $|\mbox{Normal}(10,\, 2)|$  \\
			$\kappa_{H}$     &  Hospitalisation rate   &   \num{0.1}  &  $\mbox{T.Normal}(0.1,\, 0.1)$; $[0,1]$ \\
			$\kappa_{CC}$    &  Critical care rate   &   \num{0.1} & $\mbox{T.Normal}(0.1,\, 0.1)$; $[0,1]$  \\
			$\mu_H$     &  Hospitalised death rate   &   \num{0.5} & $\mbox{T.Normal}(0.5,\, 0.5)$; $[0,1]$\\
			$\mu_{CC}$  &  Critical care death rate   &   \num{0.5} & $\mbox{T.Normal}(0.5,\, 0.5)$; $[0,1]$\\
			$\Delta_{d(t)}$  &  Reporting correction parameter for day $d(t)$  &   \num{0.14} & $\mbox{Beta} ( 10,\, 60 )$ \\
			$\Delta^D_{d(t)}$  & Monday-Friday reporting correction parameter for deaths on day $d(t)$  &  \num{1}  & $\mbox{Beta} ( 10,\, 60 )/\num{0.14}$ \\
			$\Delta^D_{d(t)}$  & Weekend reporting correction parameter for deaths on day $d(t)$  &    \num{0.1}  & \num{0.1}$\times \mbox{Beta} ( 10,\, 60 )/\num{0.14}$ \\
			$  \omega_I   $    &  Overdispersion (cases)    &   \num{100} &  $\mbox{Exp} ( 100 )$ \\
			$  \omega_H   $    &  Overdispersion (hospitalisation)    &   \num{100} &  $\mbox{Exp} ( 100 )$ \\
			$  \omega_{CC}   $    &  Overdispersion (critical care)    &   \num{100} &  $\mbox{Exp} ( 100 )$ \\
			$  \omega_D   $    &  Overdispersion (deaths)    &   \num{100} &  $\mbox{Exp} ( 100 )$ \\
			
			\midrule
			\multicolumn{1}{@{}l}{Variable}&\multicolumn{1}{l}{Description}&\multicolumn{2}{l}{Initial value}  \\
			\midrule
			$S_t $        &  {Susceptible population}       
			& \multicolumn{2}{l}{$N-I_0-E_0$} \\
			$E_t $        &  {Exposed population}
			& \multicolumn{2}{l}{$4\times I_0$} \\
			$I_t $        &  {Infected population}
			& \multicolumn{2}{l}{$I_0$} \\           
			$H_t $ &  {Hospitalised population }
			& \multicolumn{2}{l}{$0$} \\
			$C_t $ & {Intensive care population }
			& \multicolumn{2}{l}{$0$} \\
			$D_t $        &  {Dead population}
			& \multicolumn{2}{l}{$0$} \\
			$R_t $ & {Recovered population }
			& \multicolumn{2}{l}{$0$} \\
			\bottomrule
		\end{tabularx}
		\caption{Model variables and parameters, alongside their estimated prior values and distributions \cite{Gamma_dist_Appropriate_Models_For_infectious_diseases}. $N$ is the total population size  \num{5463300}, based upon the \num{2019} census data \cite{Scotland_population_data_2019}. $I_0$ is the number of cases at the first point, scaled appropriately according to equation (\ref{Observ_I}).}\label{TABLE:Model_parameters} 
	\end{table}
	
	A few comments are needed here. We have found that the model found it difficult to distinguish between the epidemic growth associated with the virus transmission and hence regulated by $\beta$ and the increase in the importation rate, $\epsilon$, and the background rate, $\phi$; subsequently, both parameters $\epsilon$ and $\phi$ have been fixed equal to 0 here. The average of the weekday reporting rates, $\Delta_{d(t)}$, is chosen so that the ratio of reported to unreported cases is 1:7. The weekend death reporting correction is set at 10\% of the Monday-Friday value from the initial observations of the reported deaths. Finally, we estimated the initial proportion of latent to infectious individuals as 4:1 to reflect the rapid initial growth of the epidemic.
	
	\subsection*{Bayesian parameter estimation and the particle filtering algorithm}
	In order to evolve the parameter values in time, and estimate their most likely values at a given time, we make use of Bayesian inference, and denote the parameter space at any given time by $\Theta_t$ and the state space by $\Sigma_t$.  
	
	The daily COVID-19 surveillance data for the Scottish population \cite{Scotland_Covid19_trends} will be denoted 
	\begin{equation*}
	\hat{\Omega}_t = ( \hat{I}^{\mbox{rep}}_t,\, \hat{H}^{\mbox{rep}}_t,\, \hat{C}^{\mbox{rep}}_t, \, \hat{D}^{\mbox{rep}}_t  ), 
	\end{equation*}
	where the hat is used to distinguish a real-world data stream, and the four terms on the right hand side correspond to the number of daily reported cases, current hospital bed occupancy, current critical care numbers, and daily deaths, respectively.
	
	Under the Bayesian approach, this data is incorporated into the likelihood function, to obtain the most likely values of the parameters within the parameter space $\Theta$ at each daily time step $t$. This results in a time series made up of the posterior distributions for both the parameters $\Theta$, and the state space $\Sigma_t$. 
	
	A statistical method called particle filtering \cite{doucet2000sequential} is used. This technique is a Monte Carlo type algorithm which assigns weights to each of the possible scenarios described by the posterior distribution. At each time step, particle filtering integrates over all potential parameter and state space combinations to calculate the likelihood that each of these configurations would occur in real life, given the known real world data from $\hat{\Omega}_t$. Each unique combination of parameter values $\Theta$ and state space $\Sigma_t$, is termed one ``particle''. A series of \num{25000} of these particles (possible parameter combinations), is generated from the prior distribution. Each particle is integrated forward in time by one timestep, and the likelihood of the new parameter/state space configuration occurring, given the data, is evaluated using a likelihood function. Each particle is then assigned an appropriate weight, depending how likely the scenario it describes is to occur in real life \cite{ong2010real}. Resampling techniques are used to avoid potential for all weights being assigned to only a small number of particles, whilst particle diversity is maintained by the use of kernel smoothing; full details of which can be found in \cite{trenkel2000fitting}. 
	
	The particle filter algorithm follows the steps below:
	\begin{enumerate}
		\item At time $t=0$, a set of $P=25,000$ \textbf{initial particles are generated} from the prior distributions outlined in Table \ref{TABLE:Model_parameters}. Each particle $p$ can be written as a vector $x_t^p = ( \Sigma_t, \Theta_t)$, with an associated weight $w_t^p$. Thus initially, we have $x_0^p = ( \Sigma_0, \Theta_0)$, and set $w_0^p = \frac{1}{P} \, \forall \, p $, such that all particle weights are initially equal.
		\item The state space is \textbf{iterated forward} in time. To achieve this, $\Sigma_{t+1}$ at time $t+1$ is drawn using Monte Carlo simulation, which involves solving equations (\ref{Gov_Eqn}) by inputting each particle's conditional distribution $x_t^p$ and weight $w_t^p$ to generate a new distribution.
		\item The particles are then \textbf{reweighted}, by first setting this new distribution to be $\tilde{x}_{t+1}^p = (\Sigma_{t+1}, \Theta_t  )$, and then calculating the corresponding likelihood contribution $\mathcal{L}_{t+1}^p =  f (\hat{\Omega}_{t+1}|\tilde{x}_{t+1}^p )$. The particle weights are then modified to be $\tilde{w}_{t+1}^p =  \mathcal{L}_{t+1}^p $, and scaled to ensure they sum to one, such that :
		\begin{equation*}
		\tilde{w}_{t+1}^p = \tilde{w}_t^p   \, / \, \Sigma_{q=1}^{P} \tilde{w}_t^q.
		\end{equation*}
		\item Next, particles are \textbf{resampled} and those whose weights are too small are rejected; typically 50-75\% are rejected in this step.
		\item \textbf{Kernel smoothing} \cite{trenkel2000fitting} is then used, to maintain particle diversity. The new set of \num{25000} particles are generated from a multivariate log-normal distribution with the average and variance calculated from the particles which survived in step 4.
		\item  Finally, steps \num{2}--\num{5} are repeated until the current time is reached, in order to \textbf{advance the system in time}. After the current time has been reached, only step \num{2} is repeated in order to obtain a forecast of the desired length.  
	\end{enumerate}
	
	\subsection*{Forecasting}
	
	The same model is used for forecasting over a period of 14 days into the future, as this is a typical horizon over which predictions are of interest to public health officials. It also represents a period when the parameters can be assumed to be constant. We used the procedure described in step \num{2} above to advance the system state over time, assuming all parameters are the same. However, to account for the level of the current testing efficiency and its variation over a week, we used equation (\ref{Observ_I}) with $Z_t$ values set as per the same weekday in the week immediately prior to the date from which the predictions begin. This is done in addition to the weekday variation captured by $\Delta_{d(t)}$ and $\Delta^D_{d(t)}$.
	
\section{Results} 
\subsection*{Real-time fitting to multiple surveillance streams}
Simulation results and forecasts from the model are presented in Figures  \ref{Figure_3}--\ref{Figure_9} below. The model is used to analyse the progress of the COVID-19 pandemic in Scotland, from \num{8}\textsuperscript{th} March--\num{30}\textsuperscript{th} November 2020, using the five streams of Scottish surveillance data: reported cases, number of tests, hospitalisation numbers, critical care numbers, and deaths, illustrated in Figure \ref{Figure_1}. We have used 25,000 particles and run the model for 268 days, with 14-day forecasts predicted for each of these days. Predictive credible intervals (50\% and 95\%; CIs) are used to quantify the uncertainty in the simulations of the system state and for the posterior parameter distributions. We also use histograms of prior and posterior distributions to assess the relative strength of the likelihood estimations. The model is implemented in \textsf{R} statistical programming language \cite{R_Project_2020} and is available on request. 

We structure the results section by the three questions posed in the Introduction.

\subsection*{Describing the outbreak}

In this section we address the question: \textit{Can the model capture the dynamics represented by the five streams of data (reported cases, reporting intensity, hospitalisation and critical care occupancy and deaths)?}

\begin{figure}[!h]
	\centering           
	\includegraphics[width=\textwidth, trim= 0cm 0cm 0cm 0cm,clip]{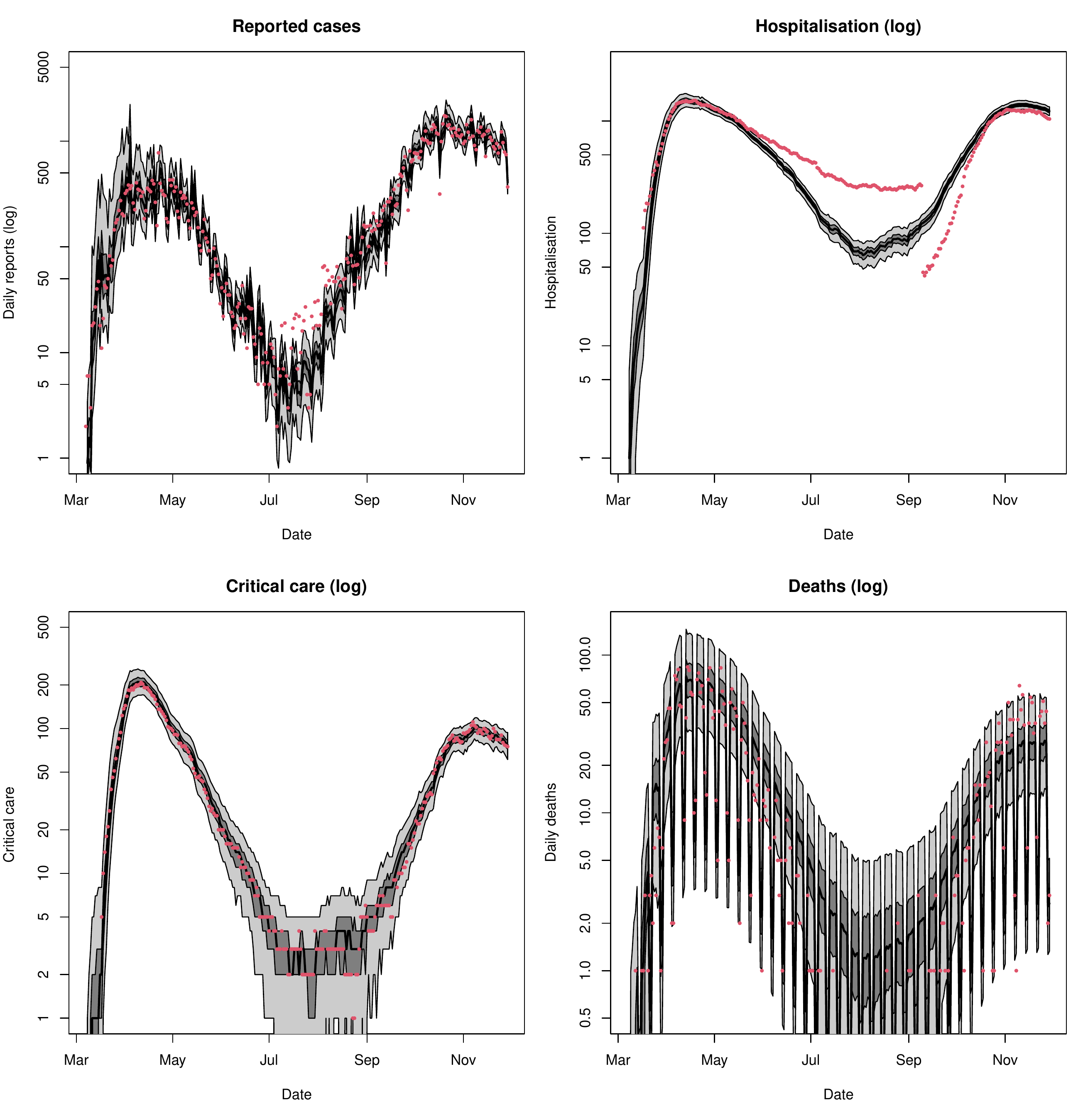} 
	\caption{Simulation results, produced by fitting the SEIR-HCD to all four Scottish surveillance data streams, from \num{8}\textsuperscript{th} March--\num{30}\textsuperscript{th} November. Median prediction is shown by the black line, alongside the corresponding $95$\% credible intervals (CI) (light grey) and $50$\% CI (dark grey). The real-world data is shown as red points. For convenience of view, the $y$-axes use a logarithmic scale. (a) Reported cases; (b) hospitalisation; (c) critical care; and (d) daily deaths.}
	\label{Figure_3}
\end{figure}

The dynamics of the reported cases, Figure \ref{Figure_3}(a), critical care numbers, Figure \ref{Figure_3}(c), and deaths, Figure \ref{Figure_3}(d), are captured by the model in excellent detail, including the slow changes over the two waves as well as short-term weekly variation. The only exception is the period of 3 weeks in July 2020 where the model is relatively slow to detect the increase in reported cases. The numbers are low in any case (between 1 and 50 cases per day) and it is possible that these largely represent the influx of new cases into Scotland. July was the period when the public was encouraged to travel for within-country holidays and the model currently does not capture such imports. The variability in the reported cases is higher during the first two months (March--May 2020) representing the uncertainty in the parameters which is explored in the next section. 

The simulated results for current hospital numbers, shown in Figure \ref{Figure_3}(b), exhibit a deviation from the data in the period from early June to September. However, it is important to note the change, mentioned previously, to the way in which the hospitalisation data was recorded from the 18\textsuperscript{th} September onwards. This removed a significant number of lingering ``false'' COVID-19 related hospitalisations from the surveillance data, i.e.\ patients who were in hospital for unrelated reasons. This resulted in overestimation of such numbers in the period prior to 15\textsuperscript{th} September. It is reassuring to see that the model output (based on all streams of information) produced consistently lower values for hospitalisation than reported in data, Figure \ref{Figure_3}(b). Moreover, the model predictions give a very similar value for hospitalisation numbers in the period shortly before 15\textsuperscript{th} September to the reduced value of the redefined Public Health Scotland data.

We also note that the predicted deaths exhibit some slight underestimation during the second wave. This could be caused by higher real case-to-report conversion rate, with the $\Delta_{d(t)}$ not being able to adjust.

\begin{figure}[!h]
	\centering           
	\includegraphics[width=0.75\textwidth, trim= 0cm 0cm 0cm 0cm,clip]{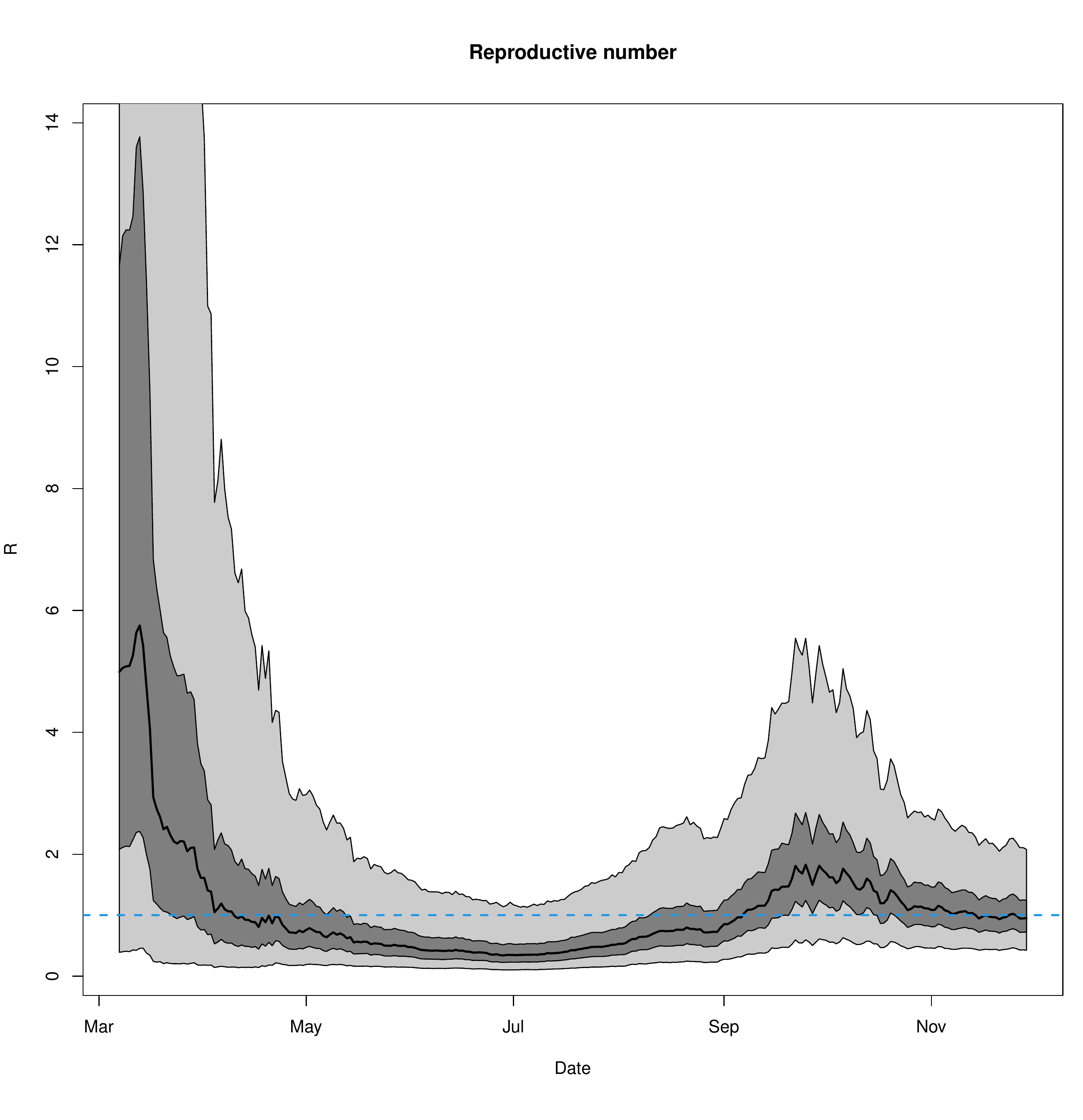} 
	\caption{Estimated effective reproductive number $R$, for COVID-19 in Scotland, for the period from \num{8}\textsuperscript{th} March--\num{30}\textsuperscript{th} November. Median prediction is shown by the black line, alongside the corresponding $95$\% CI (light grey) and $50$\% CI (dark grey). Horizontal dashed blue line corresponds to $R_0=1$. }
	\label{Figure_4}
\end{figure}

The model can also be used to calculate the effective reproductive number, $R$, using the formula

\begin{align}
R=\dfrac{\beta}{1-\exp{\left(1/\tau\right)}} ,
\end{align}
the results of which are shown in Figure \ref{Figure_4}. Although there is a large initial uncertainty, the median value starts at 5 before quickly dropping to below 1 in April 2020, following the national lockdown introduced on 23\textsuperscript{rd} March 2020. It subsequently stays below 1 (lowest median value of 0.34) until the beginning of September, although it starts to rise in July. The subsequent reintroduction of the restrictions reduces $R$ from the high median value of 1.8 back to 1 in November 2020. This behaviour and values are comparable with reported elsewhere for Scotland and globally.

\subsection*{Identifying processes and parameters}

\textit{How do the parameters change in time, reflecting the changes in the processes underlying the virus transmission and its public health implications?} We particularly identified the parameters for which the estimated values change beyond the initial ``settling-in'' period with the credible intervals narrower than for the prior and displaying long-term changes over the period of the epidemic; this group includes $\beta$, $\mu_H$ and $\mu_{CC}$.

\begin{figure}[!h]
	\centering           
	\includegraphics[width=\textwidth, trim= 0cm 0cm 0cm 0cm,clip]{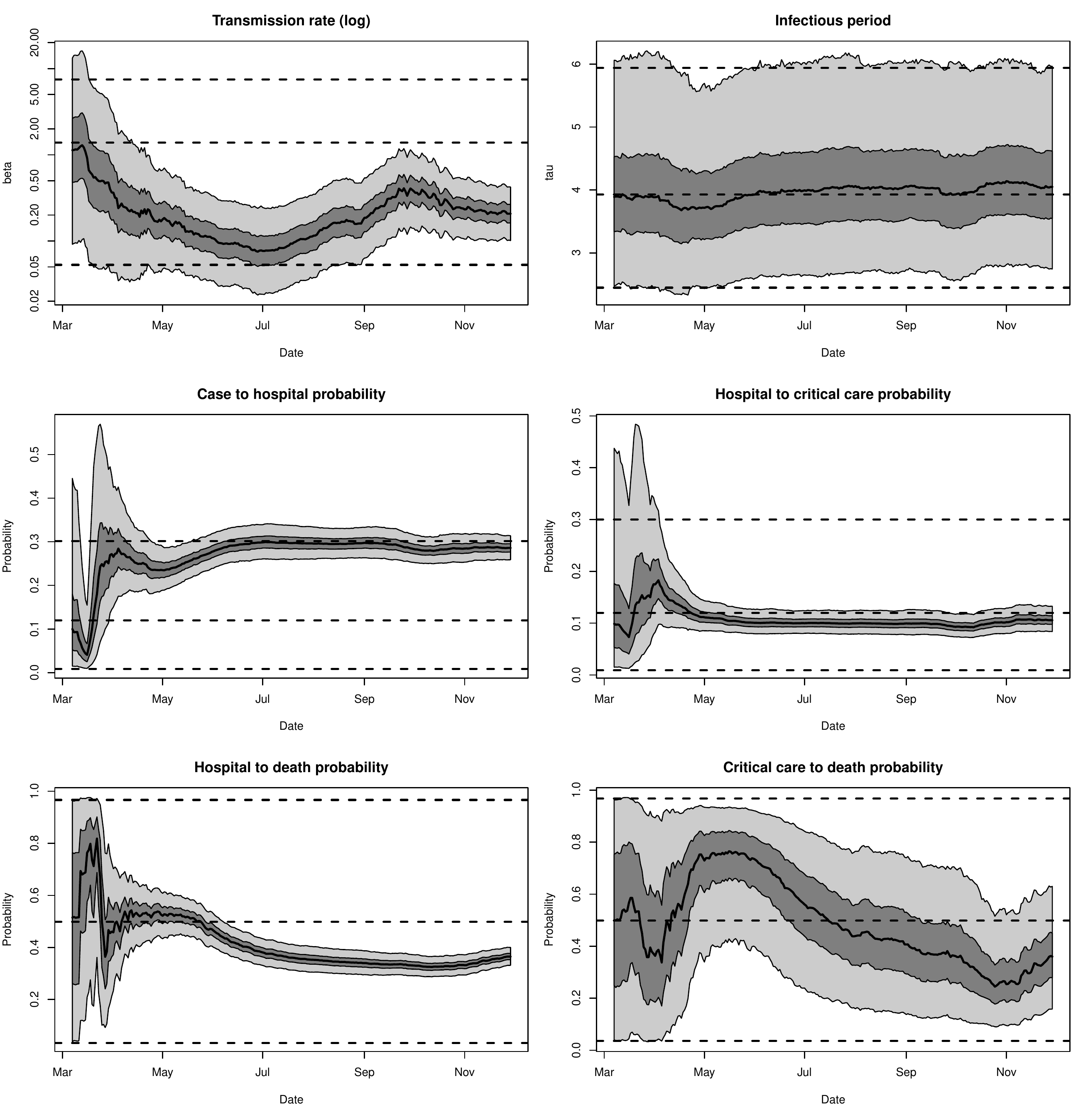} 
	\caption{Evolution of model parameters from \num{8}\textsuperscript{th} March--\num{30}\textsuperscript{th} November. Median prediction shown as black line, alongside $95$\% CI (light grey) and $50$\% CI (dark grey). Horizontal dashed lines correspond to median, and $95$\% CI values for the parameter prior estimates. (a) transmission rate, $\beta$; (b) infectious period, $\tau$; (c) case to hospital transition probability, $\kappa_H$; (d) hospital to critical care transition probability, $\kappa_C$; (e) hospital death probability, $\mu_H$; (f) critical care death probability, $\mu_{CC}$.}
	\label{Figure_5}
\end{figure}

\begin{figure}[!h]
	\centering           
	\includegraphics[width=\textwidth, trim= 0cm 0cm 0cm 0cm,clip]{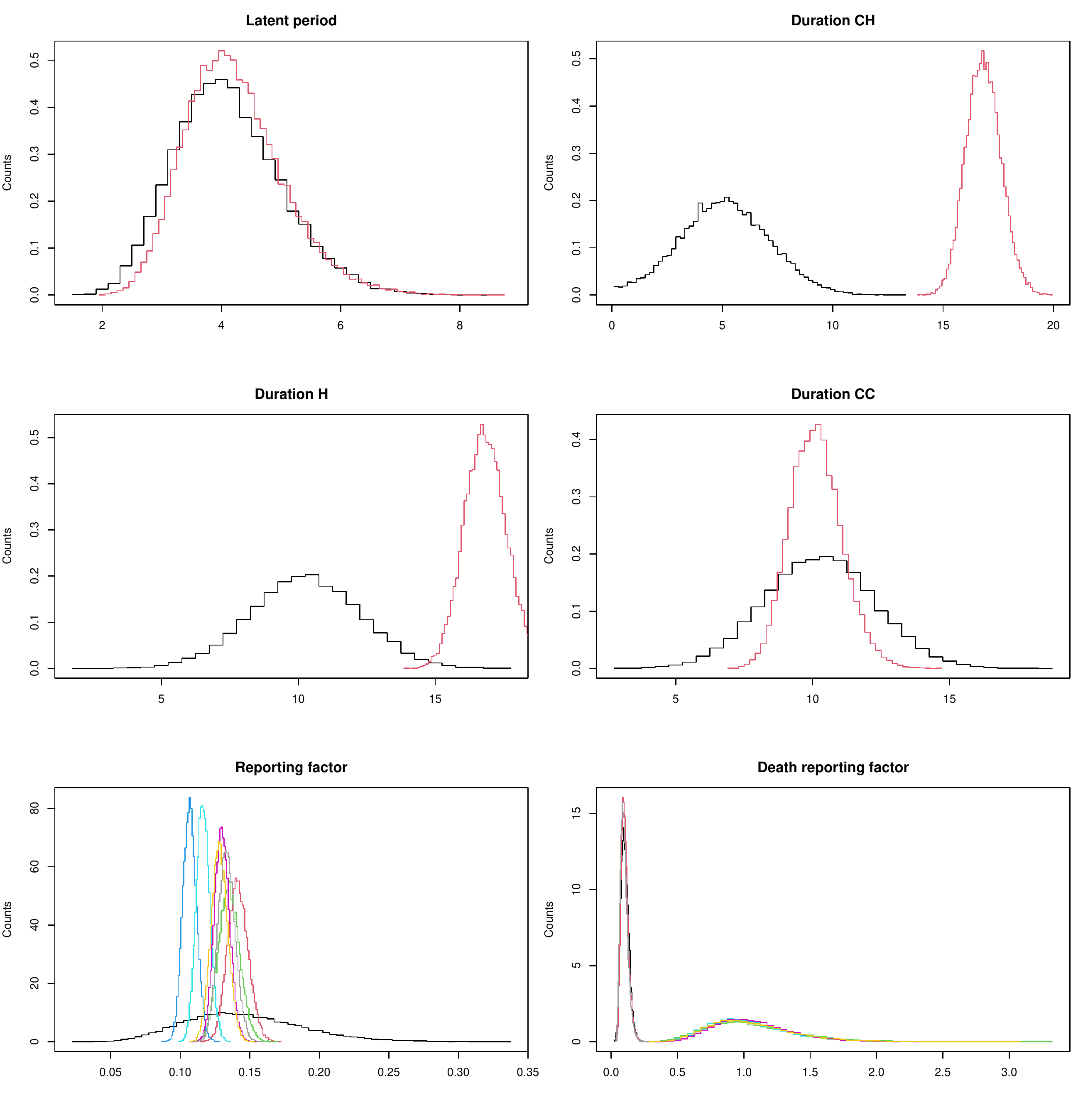} 
	\caption{Estimated prior distributions (black) and predicted posterior distributions (red in (a)--(d); multi-coloured in (e) and (f), to reflects the different weekdays, $d(t)$) for the model parameters: (a) latent period, $\alpha$; (b) duration of case to hospital, $\lambda_C$; (c) duration of hospital to recovery, $\lambda_H$; (d) duration of critical case to recovery, $\lambda_{CC}$; (e) weekday correction, $\Delta_{d(t)}$; (f) weekday death correction, $\Delta^D_{d(t)}$.}
	\label{Figure_6}
\end{figure}

The transmission rate, $\beta$ reflects the changes in the reproductive rate, Figure \ref{Figure_5}(a), starting at a value near 2~day$^{-1}$, dropping down to 0.1~day$^{-1}$ before rising again to nearly 0.5~day$^{-1}$. For most of the time, the posterior for $\beta$ is narrower than the prior (marked by the horizontal lines), suggesting a strong influence of the likelihood. In contrast, the posterior distributions for the infectious period, Figure \ref{Figure_5}(b), as well as the latent period, Figure \ref{Figure_6}(a), are indistinguishable from the prior distributions, suggesting that these parameters are ill-conditioned. 

The transition probabilities for hospitalisation, $\kappa_H$, Figure \ref{Figure_5}(c), and critical care, $\kappa_{CC}$, Figure \ref{Figure_5}(d), are well-defined and stay largely constant over the period. Although the case-to-hospital transition rate is higher (0.3) than anticipated (0.1), the probability that a case becomes hospitalised on any given day is only 2\%; the relatively high value perhaps reflecting the fact that testing was largely done at hospital and therefore most reported cases already had severe symptoms. In contrast, only 10\% of hospitalised patients need intensive care, Figure \ref{Figure_5}(d).

The dynamics of the death probability for hospitalised patients, $\mu_H$, Figure \ref{Figure_5}(e), and that for critical care patients, $\mu_{CC}$, Figure \ref{Figure_5}(d), are especially interesting. After an initial period, posteriors for both parameters become much narrower than priors, but both parameters also display long-term changes. The values are initially high (50\% for the hospital cases and 80\% for the critical care), but they become much lower in the period of July 2020 onward, before rising up again in November. Such high numbers again probably reflect the fact that hospital admissions have been limited to severe cases during the first wave, with improved care resulting in the decline of deaths from July onwards.

For other parameters we only compare the final posterior distributions with the prior distributions, Figure \ref{Figure_6}; we did not see significant long-term changes for these parameters. As mentioned above, the latent period is ill-defined, Figure \ref{Figure_6}(a), but for all the other parameters the posterior distributions are much narrower than priors. The duration parameters, $\lambda_C$, Figure \ref{Figure_6}(b); $\lambda_H$, Figure \ref{Figure_6}(c); and $\lambda_{CC}$, Figure \ref{Figure_6}(d), are well defined, but the values appear to be longer than anticipated. This is possibly the result of assuming an exponential process underlying the transitions rather than fixed periods. There appear to be only small differences between the values of $\Delta_{d(t)}$, Figure \ref{Figure_6}(e), suggesting that the reporting efficiency, $Z_t$, captures most short-term variability in the reported cases. The death correction factors, $\Delta^D_{d(t)}$ are ill-conditioned, Figure \ref{Figure_6}(f), and are determined by the priors, although the initial guess of 90\% reduction in death reports in the weekend appears to provide accurate results.

\begin{figure}[!h]
	\centering           
	\includegraphics[width=\textwidth, trim= 0cm 0cm 0cm 0cm,clip]{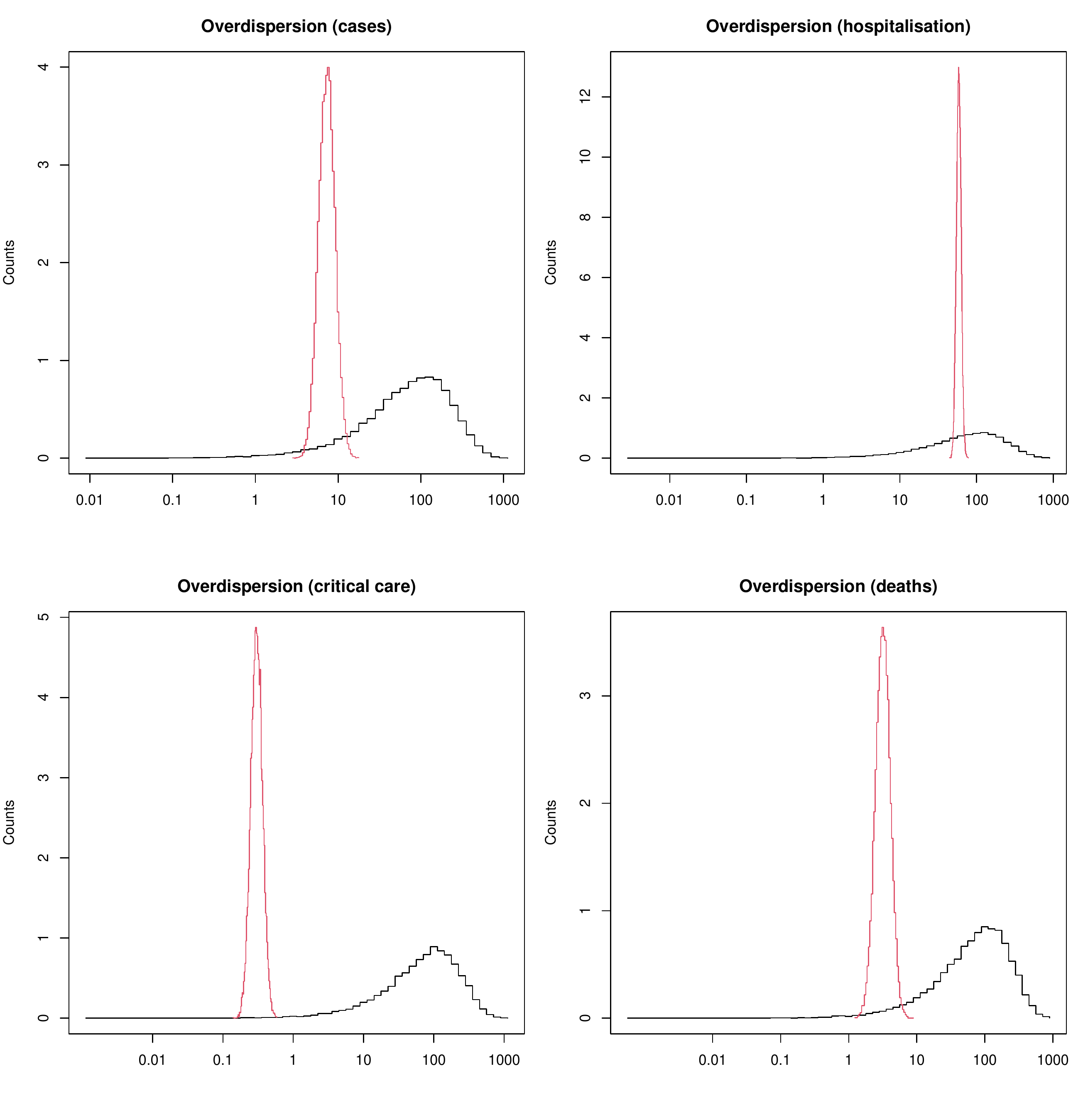} 
	\caption{ Estimated prior distributions (black) and predicted posterior distributions (red) for overdispersion terms ($\omega_I$, $\omega_H$, $\omega_{CC}$, $\omega_D$) used in the negative binomial distributions.}
	\label{Figure_7}
\end{figure}

The overdispersion parameters are well-defined and with values ranging from 0.1 for $\omega_{CC}$ to 100 for $\omega_{H}$, Figure \ref{Figure_7}. The high value of $\omega_{H}$ is related to the discrepancy between the hospitalisation model predictions and data, as shown in Figure \ref{Figure_3}.

\subsection*{Analysing short--term predictions}

\textit{Can the model be used for short-term prediction (14 days) for the key public health indicators (hospitalisation, critical care, deaths)?} In addition to the retrospective analysis of the different data streams and parameters, the model can be used to provide short-term predictions for the key variables, cases, hospitalisation, critical care, and deaths. Such predictions play a dual role. Firstly, they provide a further test of the model capability to capture the key processes. Secondly, they are essential for the public health officials, enabling them to estimate their needs. We use two cases to illustrate the potential for the model to provide such forecasts. For the initial outbreak, which was followed by a lockdown that caused the epidemic to slow, we are interested in the ability of the model to quickly respond to the limited data, even in the period when the fit is dominated by priors. We are also interested in the ability of the model to capture the effects of the lockdown. In the second period, just prior to the second wave, we are interested in the potential of the model to provide an early warning of increasing numbers and their potential to overwhelm the health service.

\begin{figure}[!hp]
	\centering           
	\includegraphics[width=\textwidth, trim= 0cm 0cm 0cm 0cm,clip]{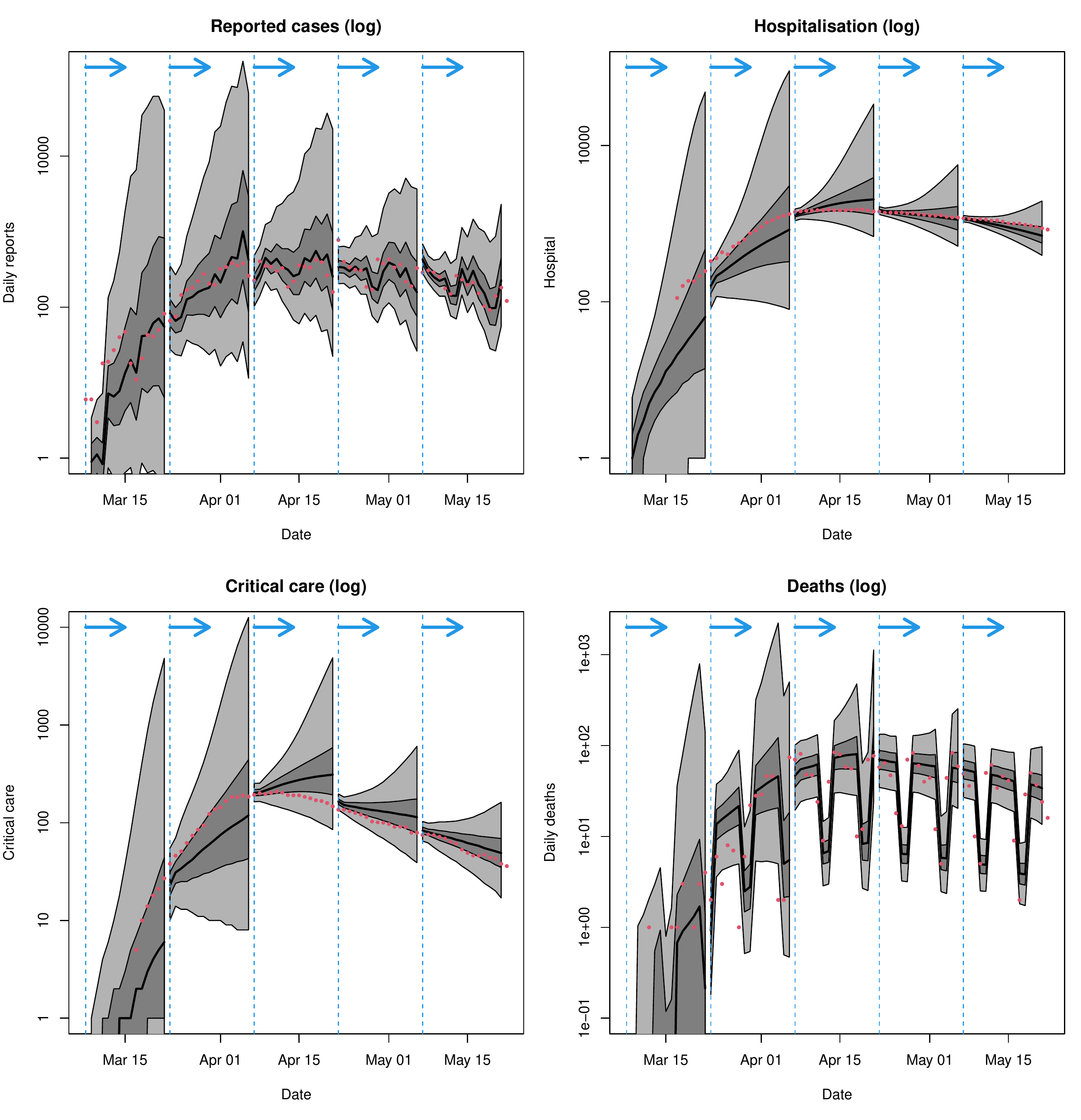} 
	\caption{First wave forecasts: Five 2-week forecasts for the second wave, from \num{8}\textsuperscript{th} March--\num{17}\textsuperscript{th} May. Vertical blue lines indicate the date each forecast was produced. Median forecast values are shown by a black line, alongside $95$\% CI (light grey) and $50$\% CI (dark grey). Real-world surveillance data is shown in red, and all $y$-axes use a log-scale. (a) Reported cases; (b) hospitalisation; (c) critical care, and (d) daily deaths. Vertical blue lines show the position of the first point in each 2-week predictions and arrows point the direction of prediction.}
	\label{Figure_8}
\end{figure}

\begin{figure}[!hp]
	\centering           
	\includegraphics[width=\textwidth, trim= 0cm 0cm 0cm 0cm,clip]{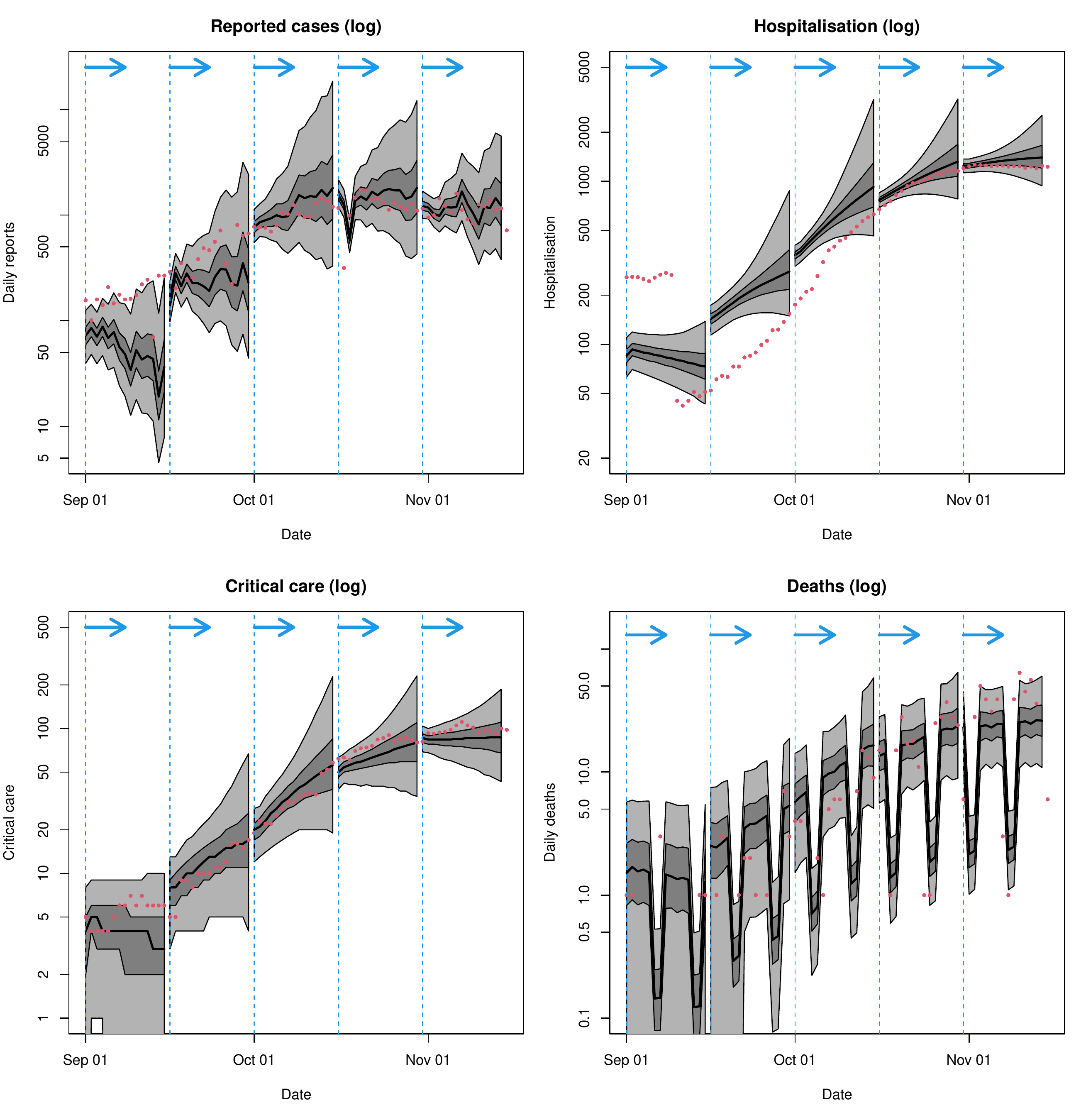} 
	\caption{Second wave forecasts: Five 2-week forecasts for the second wave, from \num{1}\textsuperscript{st} September--\num{10}\textsuperscript{th} November. Vertical blue lines indicate the date each forecast was produced. Median forecast values are shown by a black line, alongside $95$\% CI (light grey) and $50$\% CI (dark grey). Real-world surveillance data is shown in red, and all $y$-axes use a log-scale. (a) Reported cases; (b) hospitalisation; (c) critical care, and (d) daily deaths. Vertical blue lines show the position of the first point in each 2-week predictions and arrows point the direction of prediction.}
	\label{Figure_9}
\end{figure}

For each 2-week period, the predictive CIs are initially narrow but widen considerably as the time progresses, reflecting an increasing uncertainty, Figures \ref{Figure_8} and \ref{Figure_9}. In the first wave, Figure \ref{Figure_8}, the pattern is established very quickly, particularly for reported cases, hospitalisation and deaths and the model satisfactorily responds to the changes in the epidemic dynamics as lockdown reduces the spread. The short-term variation in the reported cases is also well reproduced by combining the data representing the number of reports in the previous week and the reporting correction factors, $\Delta_{d(t)}$. Forecasting the critical care occupancy is less accurate, with a slower response to the changes in the transmission around the peak, although the data stay within or just outside the 50\% predictive CIs.

Forecasting in the second wave, Figure \ref{Figure_9}, is initially hampered by two factors. Firstly, the relatively lower predictability for the number of cases in the period between June and September, Figure \ref{Figure_3}, is causing the 2-week forecast starting on 1\textsuperscript{st} September to miss the increase in cases, predicting instead the continuation of decline. As discussed above, this is most likely caused by imports of cases during summer months. Secondly, the mismatch of the data and hospital model predictions due to over-reporting of COVID-19 infections in hospitals is also causing the hospitalisation forecasting to be different from the data. However, the predictions based on the record up to 15\textsuperscript{th} September 2020 (the second prediction interval) show the correct trend for hospitalisation and excellent match with reported cases and critical care data. Forecasting improves further in later stages of the epidemic, for October and November prediction intervals shown here.
	
\section{Discussion and conclusions}

Public health responses to the pandemic rely on the ability to confidently model the epidemic spread and its consequences for hospitalisation, critical care and deaths. The models need to be parametrised based on available data. Typically, only one (or very few) stream(s) of information is used to derive the epidemiological parameters within existing models, e.g. \cite{Liu_2020March11, magal_webb_SIRU_Other_countries,zhang2020estimation}. This in turn leads to assumptions being made about the other model parameters, to which the data stream does not explicitly relate, which introduces additional uncertainty into a model. In such approaches, epidemiological parameters are found based on ``new cases'' or deaths data, whereas the other model components and parameters linked to hospitalisation and critical care occupancy are fixed. In our paper we have developed and tested the model that includes all five streams of information available through the Health Protection Scotland surveillance programme. 

We have demonstrated that the model successfully captures all data streams throughout the two pandemic waves, March--June and September-November 2020. The successful fit captures both the short-term variability in reported cases and in reported deaths, as well as to the long-term rise and fall of the epidemic and hospital/critical care pressures. The estimated values for the reproductive number, $R$, broadly agree with those obtained in other studies \cite{Scot_Gov_Modelling_the_epidemic_reports}, and shows the capacity of the model to provide early warning of the infection resurgence. The long-term changes in the hospital- and critical care-based deaths also agree with the independent studies \cite{Scot_Gov_Modelling_the_epidemic_reports}. 

The model hospitalisation predictions in the period of late May to 15\textsuperscript{th} September 2020 are consistently lower than the recorded data. Remarkably, our predictions for early September match the corrected value following the change in recording strategy of the COVID-19 hospital cases by Health Protection Scotland. Thus, the inclusion of multiple streams of data has allowed us to provide insight into the incorrect recording of one data component.

The ability of the model to provide short-term predictions is important for the public health officials. An earlier version of the model presented here have  been used successfully by Health Protection Scotland and the Scottish Government to monitor and forecast cases at national and local health board levels. The model has been incorporated into an online dashboard, accessible by all health boards to aid them in resource planning.

There are several areas in which the model needs improvement. Firstly and most importantly, the model is not aged structured and therefore is unable to capture the differential response to the infection and age--dependent death rates. It also does not account for regional and sectoral differences, including care home spread.

Secondly, some of the parameters have values that are not compatible with independent studies. This particularly applies to the length of the periods from case to hospital, hospital to critical care, and hospital and critical care to death. The disagreement is likely to be a result of simplifying assumptions made in the hospitalisation and critical care parts of the model which uses an exponentially--distributed period.

Thirdly, some parameters are ill--conditioned and their posterior distributions are not different to the priors. More work is needed on establishing what other data sets can be used to provide more information to successfully estimate their values.

Fourthly, the model currently does not include imports from outside the system which most likely have caused the underestimation of cases over the summer period. The inclusion of parameters $\phi$ and $\epsilon$ has been problematic from the parameter estimation point of view, but the process can be improved by introducing constraints on their values.

Finally, the inclusion of reporting intensity and the relationship between the reported cases and the actual number of infected individuals is currently very simplistic. The model would have benefited from the inclusion of serological data which could have provided more information on the reporting and death correction factors, $\Delta_{d(t)}$ and $\Delta^D_{d(t)}$. The reporting strategy in Scotland changed over summer 2020, switching from hospital--based testing to both hospital-- and community--based one. The model currently does not account for the differences in the type of testing, although it incorporates the testing efficiency.
	
	\section*{Acknowledgements}
	This work was partly supported by the Scottish Government Theme 2 2016-2021 Strategic Research Programme, by the Health Protection Scotland project R170923-101, and by the NERC grant NE/V00347X/1.

	\newpage
	\bibliographystyle{IEEEtran}
	\bibliography{myrefs_covid19}

\end{document}